# MACRO WITH PICO CELLS (HETNETS) SYSTEM BEHAVIOUR USING WELL-KNOWN SCHEDULING ALGORITHMS

Haider Al Kim[1], Shouman Barua[2], Pantha Ghosal[2] and Kumbesan Sandrasegaran[2]

[1]Faculty of Engineering and Information Technology, University of Technology Sydney, Australia

## ABSTRACT

*This paper demonstrates the concept of using* Heterogeneous networks (*HetNets) to improve Long Term Evolution (LTE) system by introducing the LTE Advance (LTE-A). The type of HetNets that has been chosen for this study is Macro with Pico cells. Comparing the system performance with and without Pico cells has clearly illustrated using three well-known scheduling algorithms (Proportional Fair PF, Maximum Largest Weighted Delay First MLWDF and Exponential/Proportional Fair EXP/PF). The system is judged based on throughput, Packet Loss Ratio PLR, delay and fairness.. A simulation platform called LTE-Sim has been used to collect the data and produce the paper's outcomes and graphs. The results prove that adding Pico cells enhances the overall system performance. From the simulation outcomes, the overall system performance is as follows: throughput is duplicated or tripled based on the number of users, the PLR is almost quartered, the delay is nearly reduced ten times (PF case) and changed to be a half (MLWDF/EXP cases), and the fairness stays closer to value of 1. It is considered an efficient and cost effective way to increase the throughput, coverage and reduce the latency.*

## KEYWORDS

*HetNets, LTE <E-A, Macro, Pico, Scheduling algorithms & LTE-Sim*

## 1. INTRODUCTION

In the Long Term Evolution so-called LTE, the requirements for larger coverage area, more capacity, and high data rate and low latency have led to search for cost-effective solutions to meet these demands. Hence, the development in the telecommunication networks has adopted different directions to enhance the LTE system taking into account the International Mobile Telecommunications (IMT-2000) standards that have to be satisfied [1]. Network-based technologies such as Multiple Input and Multiple Output MIMO/ advanced MIMO and Transmission/Reception Coordinated Multi-Point CoMP are LTE enhancements that introduce LTE Advance (LTE-A). Other less cost enhancements based on air interfaces are proposed, such as improving spectral efficiency involving using Heterogeneous networks (HetNets). HetNets are small and less power cells within the main macro cells with different access technologies to close up the network to the end users and increase their expectation [16].According to [2], there are two main practical HetNets classes: Macro with Femto and Macro with Pico. Femto and Pico are the small and less power cells. To save the cost, operators use the same carrier frequency in the large and small cells which, on the other hand, proposes interference challenges. Figure 1 gives the main concept of HetNets. To clarify, user in LTE is well-known as a UE.

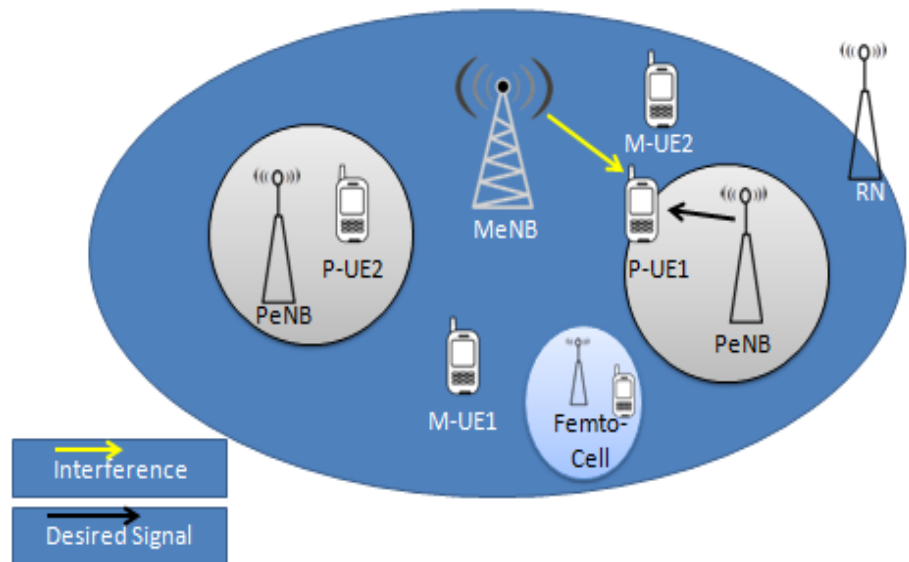

Figure.1 an example of HetNets

In LTE and LTE-A, the element that is responsible for Radio Resources Management (RRM) is enhanced Node Base station (so-called eNB). The eNB does all required management including Packet Scheduling (PS) which is the focus in the paper. PS can guarantee the agreed quality of service demands (QoS) because it is responsible for the best and effective utilizing of the affordable radio resources and in charge of data packets transmission of the users[3].

3rd Generation Partnership Project (3GPP) has left the scheduling algorithms to be vendor specific according to user's requirements and network capability. Therefore, various PS algorithms have been proposed depending on the traffic sorts and provided services. PF, MLWDF and EXP/PF algorithms [4][5][6] are used in this paper to study and compare between the system behaviours in HetNets (single Macro with 2 Pico cells) using these three types of algorithms. Scheduling algorithms ensure that QoS requirements have been met. This can be conducted by prioritizing each link between the eNB and the users, the higher priority connection the first handled in the eNB.

This paper is organized as follows. Section II discusses the downlink system model of LTE. The followed section (III) describes in more details packet scheduling algorithms, while Section IV present simulation environment. Section V shows the outcomes of the simulation. Finally, conclusion is given in Section VI.

## 2. DOWNLINK SYSTEM MODEL OF LTE

The basic element in the downlink direction of the LTE networks is called Resource Block (RB).Each UE is allocated certain number of resource blocks according to its status, the traffic type and QoS requirements. It could define the RB in both frequency domain and time domain. In the time domain, it comprises single (0.5 ms) time slot involving 7 symbols of OFDMA (orthogonal frequency division multiple access). In the frequency domain, on the other hand, it consists of twelve 15 kHz contiguous subcarriers resulting in 180 kHz as a total RB bandwidth [7].

As aforementioned before, the eNB is responsible for PS and other RRM mechanisms. The bandwidth that is used in this study is 10 MHz considering the inter-cell interference is existed.

The period that eNB performs new packet scheduling operation is the Transmission Time Interval (TTI). TTI is 1 ms that mean the users are allocated 2 contiguous radio resource blocks (2RBs). The scheduling decision in the serving eNB is made based on the uplink direction reports come from the UEs at each transmission time interval. The reports comprise the channel conditions on each RB, such as signal to noise ratio (SNR). The serving eNB uses the SNR value involved in the reports to specify the DL data rate for each served UE in each TTI. For example, how many bits per 2 contiguous RBs [8].

The data rate $dr_i(t)$ for user $i$ at $j$ sub-carrier on RB and at $t$ time can be determined by using equation (1) as proposed in [9].

$$dr_i(t) = A * B * C * D \quad (1)$$
A = $nbits_{i,j}(t)/symbol$
B = $nsymbols/slot$
C = $nslots/TTI$
D = $nsc/RB$rgg

The number of bits per symbol is "A". The number of symbols per slot is "B". While "C" represents how many slots per TTI, "D" clarifies how many sub-carriers per RB. Table 1 summarizes the mapping between SNR values and their associated data rates.

Table 1. Mapping between instantaneous downlink SNR and data rate

| Minimum SNR Level (dB) | Modulation and coding | Data Rate (Kbps) |
|---|---|---|
| 1.7 | QPSK (1/2) | 168 |
| 3.7 | QPSK (2/3) | 224 |
| 4.5 | QPSK (3/4) | 252 |
| 7.2 | 16 QAM (1/2) | 336 |
| 9.5 | 16 QAM (2/3) | 448 |
| 10.7 | 16 QAM (3/4) | 504 |
| 14.8 | 64 QAM (2/3) | 672 |
| 16.1 | 64 QAM (3/4) | 756 |

Upon the packets reach the eNB, they are buffered in eNB in a specific container allocated for each active UE. Moreover, the buffered packets are assigned a time stamp to ensure that they will be scheduled or dropped before the scheduling time interval is expired, and then using First-In-First-Out (FIFO) method they are transmitted to the users in the downlink direction. To explain the scheduling operation, PS manager (is a part of eNB functionalities) at each TTI priorities and classifies the arriving users' packets according to preconfigured scheduling algorithm.

Scheduling decision is made based on different scheduling criteria that have been used in various algorithms. For example channel condition, service type, Head-of-Line (HOL) packet delay, buffer status, and so on so forth. One or more RBs could be allocated to the selected user for transmission with the highest priority. Figure 2 shows the packet scheduler in the downlink direction at eNB.

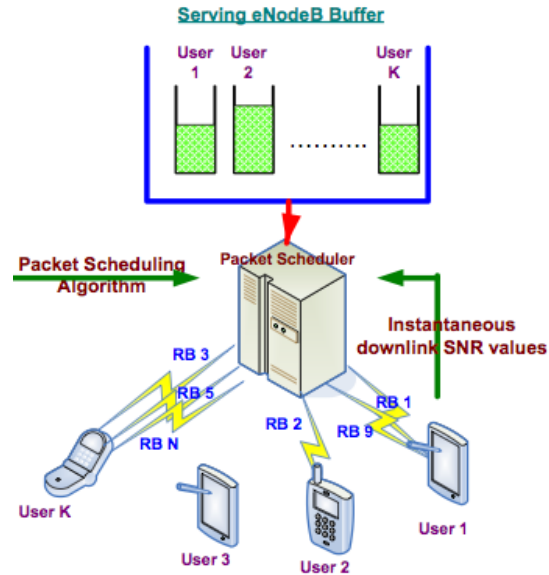

Figure.2 Downlink Packet Scheduler of the 3GPP LTE System [10]

## 3. PACKET SCHEDULLING ALGORITHMS

The efficient radio resource utilization and ensuring fairness among connected users, as well as satisfying QoS requirements, are the main purposes of using PS algorithms [11].The PS algorithms that have been used in this study are : Proportional Fair (PF) algorithm, Maximum-Largest Weighted Delay First (MLWDF or ML) and the Exponential/Proportional Fair (EXP/PF or EXP) algorithm. It should be noted that these algorithms are used.

### 3.1. Proportional Fair (PF) Algorithm

For non-real time traffic, the PF was proposed which is used in a Code Division Multiple Access-High Data Rate (CDMA-HDR) system in order to support Non-Real Time (NRT) traffic. In this algorithm, the trade-off between fairness among users and the total system throughput is presented.

This is, before allocating RBs, it considers the conditions of the channel and the past data rate. Any scheduled user in PF algorithm is assigned radio resources if it maximizes the metric *k* that calculated as the ratio of reachable data rate $r_i(t)$ of user *i* at time *t* and average data rate $R_i(t)$ of the same user at the same time interval *t*:

$$k = \arg\max \frac{r_i(t)}{R_i(t)} \qquad (2)$$

where;

$$R_i(t) = \left(1 - \frac{1}{t_c}\right) * R_i(t-1) + \frac{1}{t_c} * r_i(t-1) \qquad (3)$$

$t_c$ is the window size used to update the past data rates values in which the PF algorithm maximizes the fairness and throughput for any scheduled user. Unless user *i* is selected for transmission at $(t-1)$, $r_i(t-1) = 0$.

### 3.2. Maximum Largest Weighted Delay First (MLWDF) Algorithm

If the traffic is a Real Time (RT), the MLWDF is introduced which is used in CDMA-HDR system in order to support RT data users [11]. It is more complex algorithms compare with PF and is used in different QoS user's requirements. This is because it takes into account variations of the channel when assigning RBs. Moreover, if a video traffic scenario, it takes into consideration time delay. Any user in MLWDF is granted RBs if it maximizes the equation below:

$$k = \arg\max a_i W_i(t) \frac{r_i(t)}{R_i(t)} \qquad (4)$$

where;

$$a_i = -\frac{(\log \delta_i)}{\tau_i} \qquad (5)$$

where $W_i(t)$ is a difference in time between current and arrival times of the packet that known as the Head Of Line (HOL) packet delay of user *i* at time *t*.

Similarly to PF equation, while the achievable data rate of user *i* at time *t* is $r_i(t)$, the average data rate of the same user at the same time interval *t* is $R_i(t)$. $\tau_i$ and $\delta_i$ are the delay threshold for a packet of user *i* and the maximum HOL packet delay probability of user *i* respectively. The later is considered to exceed the delay threshold of user *i*.

### 3.3. Exponential/Proportional Fair (EXP/PF) Algorithm

Since PF is not designed for multimedia applications (only for NRT traffic), an enhanced PF called EXP/PF algorithm was proposed in the Adaptive Modulation and Coding and Time Division Multiplexing (AMC/TDM) systems. The EXP/PF algorithm is designed for NRT service or RT service (different sorts of services). The $k$ metric is used for both RT nad Non-RT in which RBs are assigned to users based on $k$.

$$k = \arg\max \begin{cases} \exp\left(\frac{a_i W_i(t) - a\overline{W(t)}}{1+\sqrt{a\overline{W(t)}}}\right) \frac{r_i(t)}{R_i(t)} & i \epsilon RT \\ \frac{w(t)}{M(t)} \frac{r_i(t)}{R_i(t)} & i \epsilon NRT \end{cases} \qquad (6)$$

where,

$$\overline{aw(t)} = \frac{1}{N_{RT}} \sum_{i \epsilon RT} a_i W_i(t) \qquad (7)$$

$$w(t) = \begin{cases} w(t-1) - \varepsilon & W_{max} > \tau_{max} \\ w(t-1) + \frac{\varepsilon}{k} & W_{max} < \tau_{max} \end{cases} \qquad (8)$$

where the average number of packets at the buffer of the eNB at time *t* is represented by $M(t)$, $k$ and $\varepsilon$ in equation (8) are constants, $W_i(t)$ is explained in MLWDF, $W_{max}$ is the HOL packets delay of RT service and $\tau_{max}$ is the maximum delay of RT service users. The EXP/PF differentiates between RT and NRT by prioritizing RT traffic users over the NRT traffic users if their HOL values are reaching the delay threshold.

## 4. SIMULATION ENVIRONMENT

LTE-Sim simulator is used in this paper to do the entire analysis and study [12]. The most recent version of LTE-Sim (version 5) has not involved yet any code regarding the HetNets type (Macro with Pico cells). The developed code used in this paper could be considered as an enhancement of the released LTE-Sim versions. However, LTE-Sim has a detailed code (or what authors are named it: scenario) which can be used to simulate and examine HetNets type (Macro with Femto). Our paper is based on a scenario of a single Macro cell with 2 small Pico cells that are reduced their powers. More Picos can be added to the system, and enhanced system behaviour will be presented. However, according to [2], while the number of Pico cells is increased, more inter-cell interference is experienced since the same carrier frequency is used in each cell (Macro and Picos).

Figure 3 shows the entire system that is used in this paper: Macro cell of 1 km and 2 Pico cells of 0.1 km located on the Macro edge. This design is chosen to analog a real system aimed to cover larger area and more users, especially the users in the cell edge where they suffer from lack of connectivity with Macro cell. The inter-cell interference is modeled. Video and VoIP traffic are used to represent user's data. Each user has 50 % Video traffic and 50% VoIP flows.

Handover is activated. Each cell starts a certain number of users. Non-uniform user distribution within the cells is applied and 3km/h constant speed is utilized as the mobility user speed. In addition, the 3GPP urban Macro cell propagation loss model has been implemented including path-loss, penetration loss, multi-path loss and shadow fading which are summarized below [13]:

- Pathloss: $128.1 + 37.6 \log_{10}(d)$, *d* refers the distance between the eNB and the user in kilometers.
- Penetration loss: 10 *dB*
- Multipath loss: using one of the well-known methods called *Jakes model*
- Shadow fading loss (recently it could be used as a gain in LTE-A): *log-normal distribution*
  - *Mean value of 0 dB.*
  - *Standard deviation of 10 dB.*

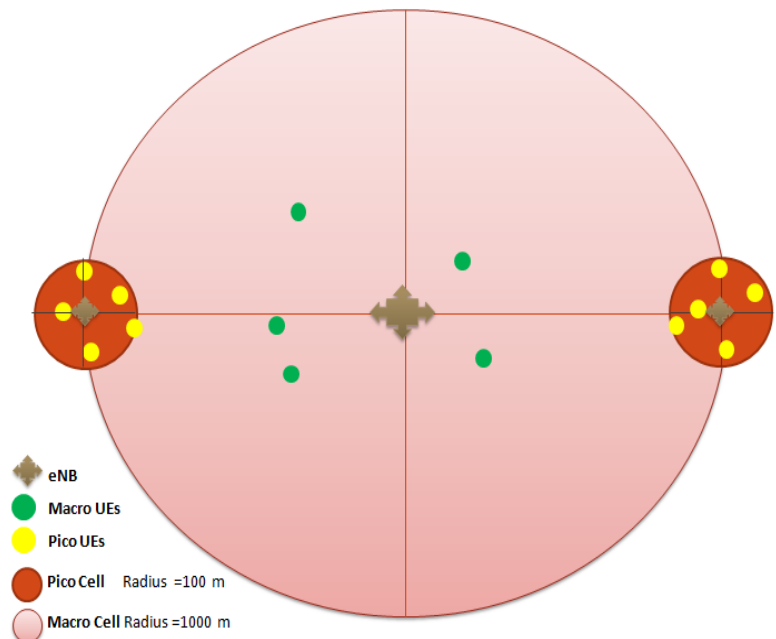

Figure.3 Applied HetNets (Macro with 2 Picos)

Packets throughput (see equation 9), Packet Loss Ratio (PLR) as shown in equation 10, packet delay (latency) and fairness index (equation 11) are the concepts used in the aforementioned algorithms to evaluate the system performance. Jain's method is applied to implement fairness among users [14]. According to [1], fairness should reach the value of 1 to be considered as a fair algorithm that sharing the resources suitably among users. It can be calculated as value 1 minus the value of the difference between the maximum and minimum size of transmitted packets of the most and least scheduled users. Equation (11) calculates the fairness value.

$$throughput = \frac{1}{T} \sum_{i=1}^{K} \sum_{t=1}^{T} ptransmit_i(t) \qquad (9)$$

$$PLR = \frac{\sum_{i=1}^{K} \sum_{t=1}^{T} pdiscard_i(t)}{\sum_{i=1}^{K} \sum_{t=1}^{T} psize_i(t)} \qquad (10)$$

$$fairness = 1 - \frac{ptotaltransmit_{max} - ptotaltransmit_{min}}{\sum_{i=1}^{K} \sum_{t=1}^{T} psize_i(t)} \qquad (11)$$

Obviously, while $ptransmit_i(t)$ is the size of transmitted packets, $pdiscard_i(t)$ is the size discarded or lost packets during the connection. $psize_i$ is the summation of all arrived packets that are buffered into serving eNB [1].

The aforementioned total size of transmitted packets of the best served UE and the worse served UE are represented in equation (11) as $ptotaltransmit_{max}$ and $ptotaltransmit_{min}$.

Table 2 shows the entire system simulation parameters [1].

Table 2. LTE system simulation parameters

| Parameters | |
|---|---|
| Simulation time | 30 s |
| Flow duration | 20 s |
| Slot duration | 0.5 ms |
| TTI | 1 ms |
| Number of OFDM symbols/slot | 7 |
| Macro cell radius | 1 km |
| Macro eNB Power | 49 dBm |
| Pico cell radius | 0.1 km |
| Pico eNB Power | 30 dBm |
| User speed | 3 km/h |
| VoIP bit rate | 8.4 kbps |
| Video bit rate | 242 kbps |
| Frame structure type | FDD |
| Bandwidth | 10 MHz |
| Number of RBs | 50 |
| Number of subcarriers | 600 |
| Number of subcarriers/RB | 12 |
| Subcarrier spacing | 15 KHz |

In order to get better results and to confirm the outcomes, five simulations have been conducted for each algorithm (PF, MLWDF and EXP) in each point of users (10, 20, 30, 40, 50, 60, 70 and 80). This yields 120 simulations outcomes. The average values have been taken to draw the simulation graphs at each point of users.

## 5. SIMULATION RESULTS

The average overall system throughput is shown in figure 4. Comparing the throughput for "single Macro cell" for the same simulation parameters as shown in figure 5, the pico cells in the scenario "Macro with 2 Picos" boost the throughput by adding gain that shown as an overall system throughput increment for the same number of users. For instance, at 40 users using MLWDF, the throughput is 25 Mbps for the scenario with 2 Picos while the Macro scenario is only 9.3 Mbps. This is almost a duple value. Further points show duple and triple throughput values in the scenario of 2 Picos. However, the gain will reach a saturation level where no more gain could be shown due to the fact of limited radio resources availability while more users are added to the system. Although MLWDF and EXP have almost similar behaviour in both scenarios, a higher throughput is shown in the 2 Pico case using both algorithms. It could note that PF algorithm as shown figure 5 behaves better than the scenario of single Macro cell. PF is developed for NRT traffic, but the simulation is for Video flows (RT traffic); hence, the other simulated algorithms outperform PF.

PLR shown in the figure 6 according to [15] is the packet loss ratio for a single Macro cell. While the system is charged with more than 20 users, the PLR is increased for all experienced algorithms taking into consideration that the PF is the worst case with the video traffic. Adding two Picos to the previous system to create "Macro with 2 Picos" scenario enhances the PLR while maintaining similar system behavior for all algorithms. Approximately, the PLR in Macro with 2 Picos case is reduced to be a quarter of PLR value of single Macro cell scenario. For example, at 70 users, MLWDF has 0.1 PLR value while for the same number of users MLWDF has 0.5 PLR value in the single Macro scenario. Comparing between scheduling schemes, the worst case is the PF algorithm in both cases. Figure 7 illustrates PLR for Macro with 2 Picos.

According to [15] and as shown in figure 8, the delay in single Macro cell scenario is close to be constant for PF, MLWDF and EXP/PF with value less than 5 ms while it suffers from rapid increasing after 40 users for PF algorithm. If two Pico cells are added to the aforementioned system, a similar performance is shown, but the delay value is decreased. In addition, the threshold of PF is shifted at 60 users instead of 40 users in the single Macro case. To compare MLWDF and EXP/PF in both scenarios, a certain point in figures 8 and 9 could be explained. For example at 60 users, in a single Macro cell the delay value is 50 ms while in the Macro with 2 Picos the value is 20 ms. As a consequence, for MLWDF and EXP/PF, the delay value with two Picos is approximately half the delay value without Pico cells. One of the purposes of HetNets is to enhance the latency, and this is shown in a practical simulation illustrated in figure 9. However, the delay shows lower values (nearly 10 times lower) in the scenario of single cell with 2 Picos using PF scheme.

When the number of users increases in single Macro cell more than 30, the fairness index of all simulated algorithms is deviated down of the value "1". At 40 users, PF shows further deviation close to value 0.8 compare with other algorithms which they are around 0.9. The fairness index behaves similarly in the scenario of Macro with 2 Picos as shown in figure 11. However, the PF shows a minor different in which at 50 users it starts to decline to get the value 0.8.

## 5.1. Throughput

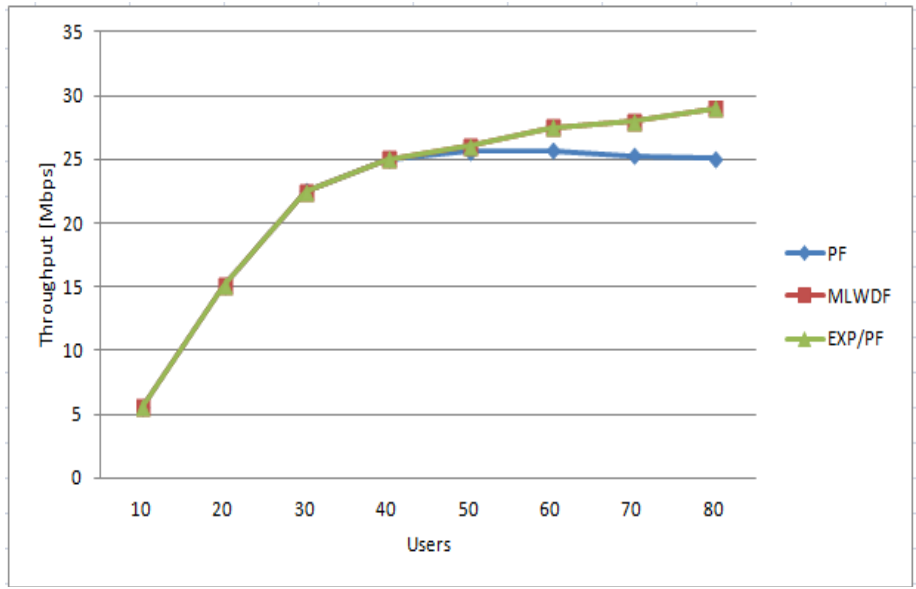

Figure.4 Average System Throughput (Macro with 2 Picos)

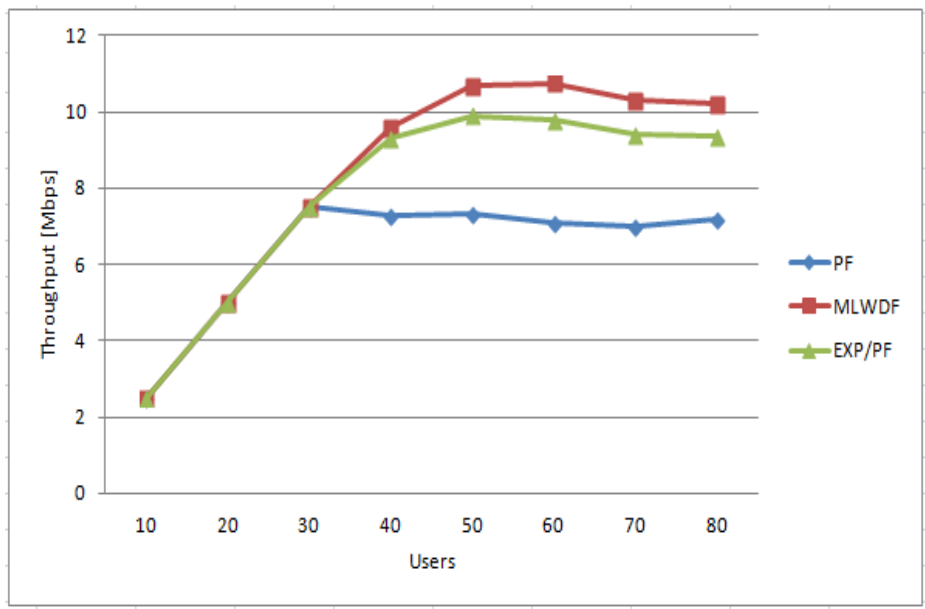

Figure.5 Average System Throughput (single Macro cell)

## 5.2. Packet Loss Ratio (PLR)

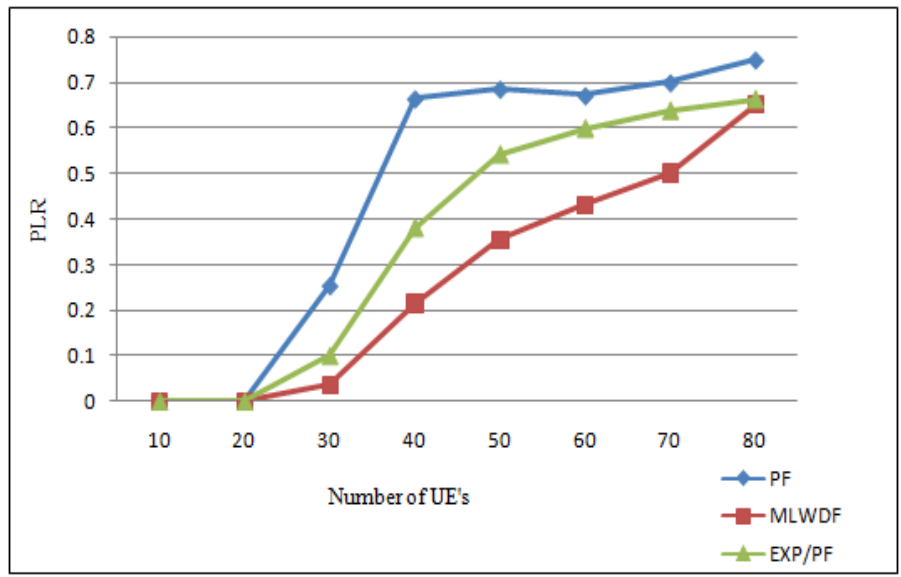

Figure.6 PLR of Video Flows (single Macro cell) [15]

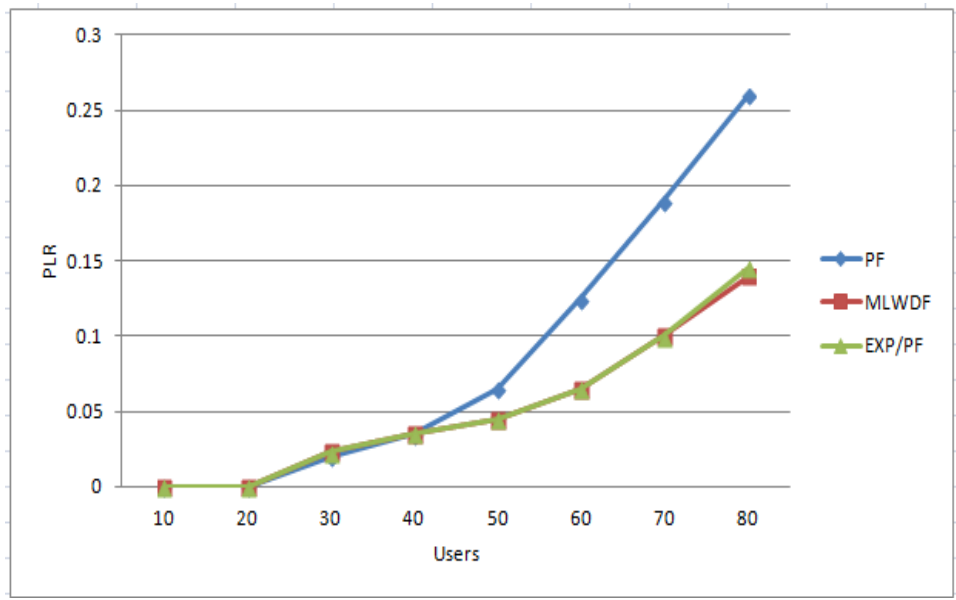

Figure.7 PLR of Video Flows (Macro with 2 Picos)

## 5.3. Delay

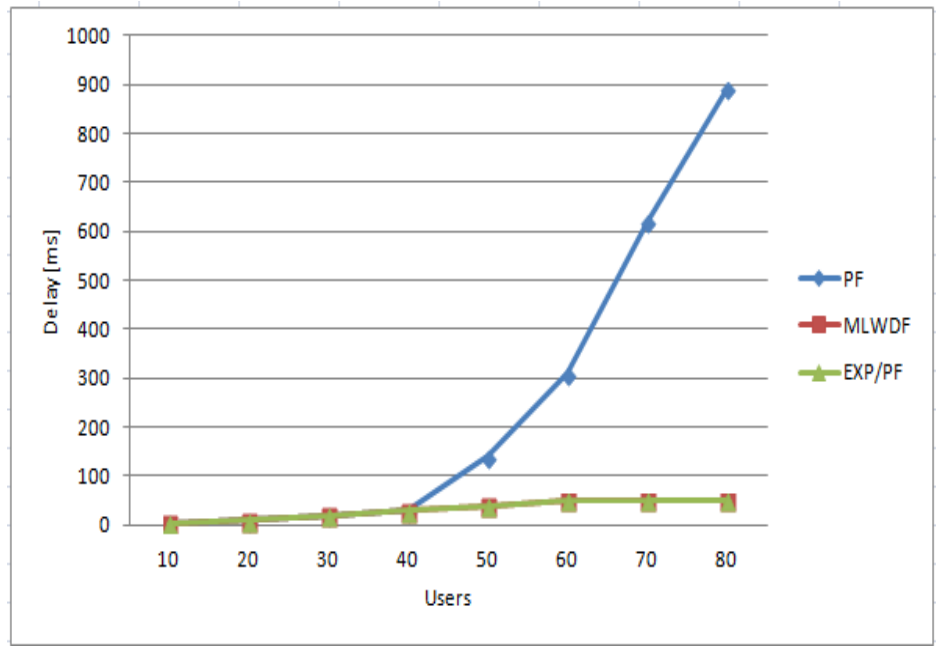

Figure.8 Packet Delay of Video Flows (single Macro cell)

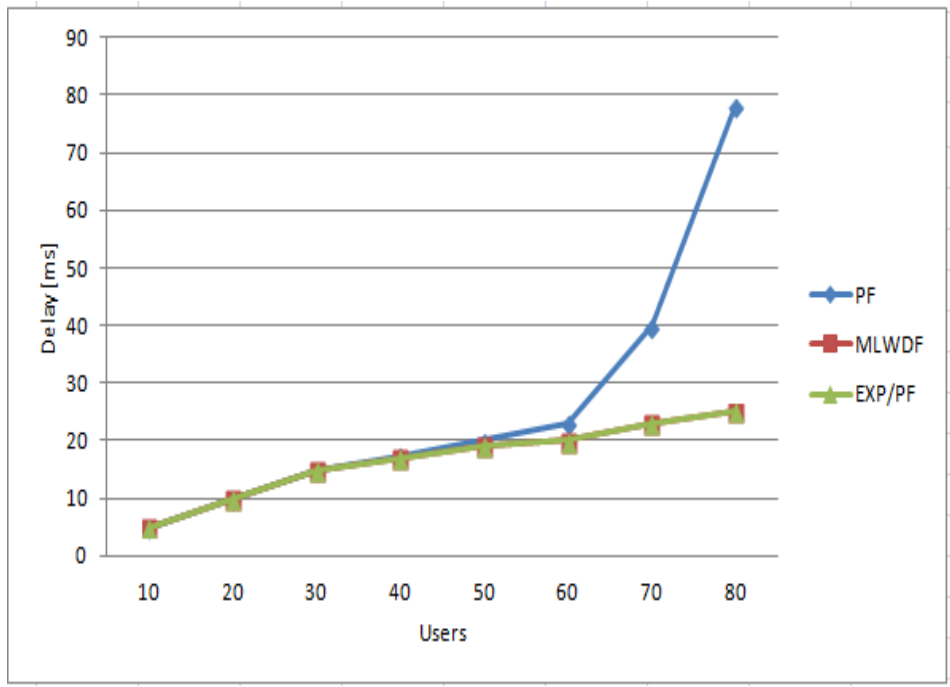

Figure.9 Packet Delay of Video Flows (Macro with 2 Picos)

**5.4. Fairness Index**

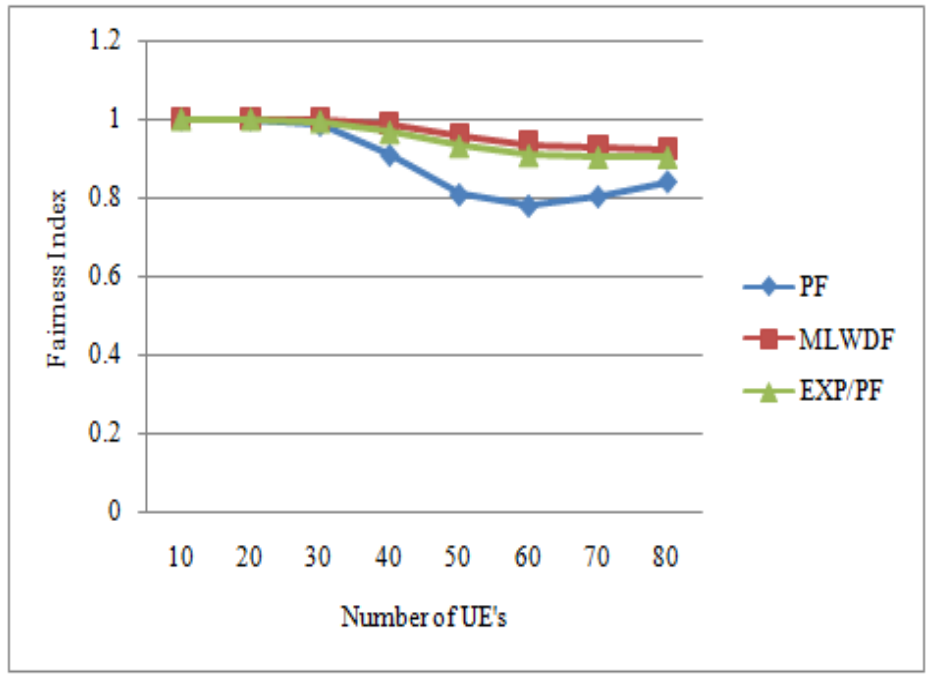

Figure.10 Fairness Index of Video Flows [15]

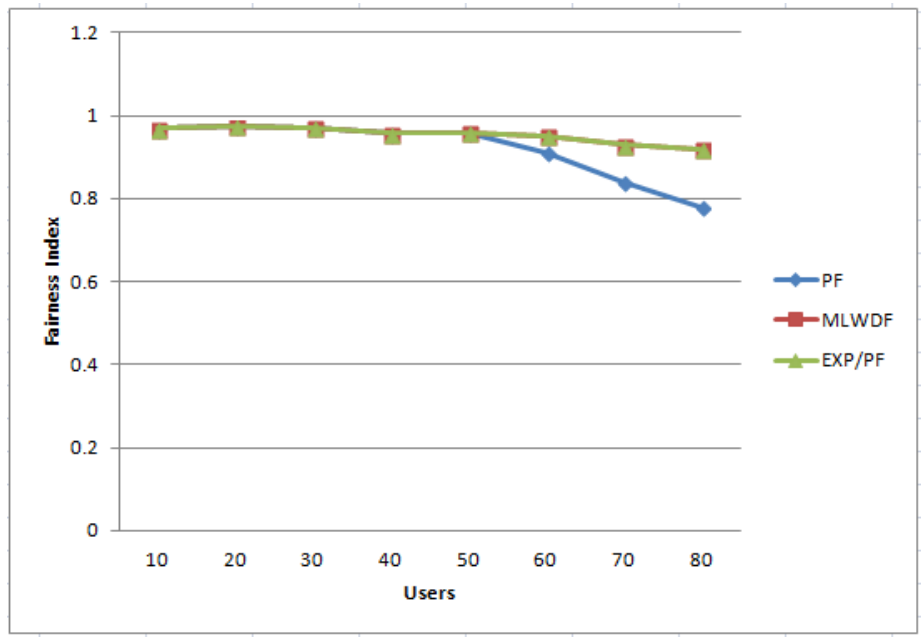

Figure.11 Fairness Index of Video Flows Macro with 2 Picos

## 6. CONCLUSION

This paper investigates scheduling algorithms that are developed to enhance the LTE network performance by sharing radio resources fairly among users utilizing all available resources. These algorithms depend on traffic class and number of users, hence; different outcomes are presented for each algorithm. To further boost the overall system performance, this study uses heterogeneous networks concept by adding small cells (2 Pico cells). This enhancement is experienced through a throughput, PLR, delay and fairness. In the throughput the system gains more data rate while in PLR the system suffers less packet loss values. Moreover, delay is decreased and fairness stays similar. Approximately from the simulation outcomes, the overall system performance is as follows: throughput is duplicated or nearly tripled relaying on the number of users, the PLR is almost quartered, the delay is reduced 10 times (PF case) and changed to be a half value (MLWDF/EXP cases), and the fairness stays closer to value of 1. As a number of small cells increases, the system is expected to be more enhanced till a saturation state is reached. The reason behind that is the inter-cell interference will limit the performance since the same carrier frequency is used in all system's cells. Focusing on macro with 2 Pico cells scenario, MLWDF shows the best performance for video flows followed by EXP/PF. Further enhancement can be applied in future papers such as almost blank subframes (ABS), enhanced inter-cell interference cancelation (eICIC) and cell range extension CRE concepts.

## REFERENCES

[1]     H. A. M. Ramli, R. Basukala, K. Sandrasegaran, and R. Patachaianand, "Performance of well known packet scheduling algorithms in the downlink 3GPP LTE system," in Communications (MICC), 2009 IEEE 9th Malaysia International Conference on, 2009, pp. 815-820.

[2]     Seung June Yi, S.C., Young Dae Lee, Sung Jun Park, Sung Hoon Jung 2012, Radio Protocols for LTE and LTE-Advanced.

[3]     [1]     B. Liu, H. Tian, and L. Xu, "An efficient downlink packet scheduling algorithm for real time traffics in LTE systems," in Consumer Communications and Networking Conference (CCNC), 2013 IEEE, 2013, pp. 364-369.

[4]     A. Jalali, R. Padovani, and R. Pankaj, "Data throughput of CDMA-HDR a high efficiency-high data rate personal communication wireless system," in Vehicular Technology Conference Proceedings, 2000. VTC 2000-Spring Tokyo. 2000 IEEE 51st, 2000, pp. 1854-1858.

[5]     M. Andrews, K. Kumaran, K. Ramanan, A. Stolyar, P. Whiting, and R. Vijayakumar, "Providing quality of service over a shared wireless link," Communications Magazine, IEEE, vol. 39, pp. 150-154, 2001.

[6]     J.-H. Rhee, J. M. Holtzman, and D. K. Kim, "Performance analysis of the adaptive EXP/PF channel scheduler in an AMC/TDM system," Communications Letters, IEEE, vol. 8, pp. 497-499, 2004.

[7]     J. Zyren and W. McCoy, "Overview of the 3GPP long term evolution physical layer," Freescale Semiconductor, Inc., white paper, 2007.

[8]     B. Riyaj, M. R. H. Adibah, and S. Kumbesan, "Performance analysis of EXP/PF and M-LWDF in downlink 3GPP LTE system," 2009.

[9]     X. Qiu and K. Chawla, "On the performance of adaptive modulation in cellular systems," Communications, IEEE Transactions on, vol. 47, pp. 884-895, 1999.

[10]    S. C. Nguyen, K. Sandrasegaran, and F. M. J. Madani, "Modeling and simulation of packet scheduling in the downlink LTE-advanced," in Communications (APCC), 2011 17th Asia-Pacific Conference on, 2011, pp. 53-57.

[11]    A. Alfayly, I.-H. Mkwawa, L. Sun, and E. Ifeachor, "QoE-based performance evaluation of scheduling algorithms over LTE," in Globecom Workshops (GC Wkshps), 2012 IEEE, 2012, pp. 1362-1366.

[12]    G. Piro, L. A. Grieco, G. Boggia, F. Capozzi, and P. Camarda, "Simulating LTE cellular systems: an open-source framework," Vehicular Technology, IEEE Transactions on, vol. 60, pp. 498-513, 2011.

International Journal of Wireless & Mobile Networks (IJWMN) Vol. 6, No. 5, October 2014

[13]   M. Iturralde, T. Ali Yahiya, A. Wei, and A. Beylot, "Resource allocation using shapley value in LTE networks," in Personal Indoor and Mobile Radio Communications (PIMRC), 2011 IEEE 22nd International Symposium on, 2011, pp. 31-35.
[14]   R. Jain, D.-M. Chiu, and W. R. Hawe, A quantitative measure of fairness and discrimination for resource allocation in shared computer system: Eastern Research Laboratory, Digital Equipment Corporation, 1984.
[15]   AL-Jaradat, Huthaifa 2013, 'On the Performance of PF, MLWDF and EXP/PF algorithms in LTE'.
[16]   Holma H, Toskala A 2012, "LTE-Advanced 3GPP Solution for IMT-Advanced".

**Authors**

**Haider Al Kim** got the B.Sc. in Information and Communication Engineering from Al-khwarizmi Engineering College, University of Baghdad, Baghdad, Iraq in 2008. He pursues his Master degree in Telecommunication Networks from University of Technology Sydney (UTS), Sydney, Australia in 2014 under the supervision A. Prof. Kumbesan Sandrasegaran. Working and research areas are Wireless Telecommunication, Mobile Network, Network Management, Network Design and Implementation and Data Analysis and Monitoring .He is senior network engineer with more than 5 years work experience in networks and telecommunication industry at University of Kufa, Iraq. He is also a Cisco Certificate holder (ID: CSCO11773718) and Cisco instructor at Al-Mansour College, Baghdad, Iraq in 2010-2011. Alcatel-Lucent SAM certification holder, Alcatel University, Sydney Australia 2013.

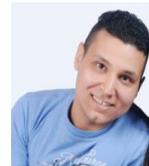

**Shouman Barua** is a PhD research scholar at the University of Technology, Sydney. He received his BSc in Electrical and Electronic Engineering from Chittagong University of Engineering and Technology, Bangladesh and MSc in Information and Communication Engineering from Technische Universität Darmstadt (Technical University of Darmstadt), Germany in 2006 and 2014 respectively. He holds also more than five years extensive working experience in telecommunication sector in various roles including network planning and operation.

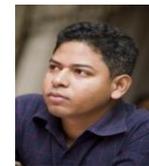

**Pantha Ghosal** is a Graduate Research Assistant at University of Technology, Sydney. Prior to this, he completed B.Sc in Electrical and Electronic Engineering from Rajshahi University of Engineering & Technology, Bangladesh in 2007. He is an expert of Telecommunication network design and holds more than 7 years of working experience in this area.

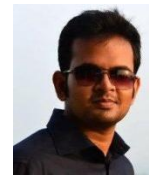

**Dr Kumbesan Sandrasegaran** is an Associate Professor at UTS and Centre for Real-Time Information Networks (CRIN). He holds a PhD in Electrical Engineering from McGill University (Canada)(1994), a Master of Science Degree in Telecommunication Engineering from Essex University (1988) and a Bachelor of Science (Honours) Degree in Electrical Engineering (First Class) (1985).  His current research work focuses on two main areas (a) radio resource management in mobile networks, (b) engineering of remote monitoring systems for novel applications with industry through the use of embedded systems, sensors and communications systems. He has published over 100 refereed publications and 20 consultancy reports spanning telecommunication and computing systems.

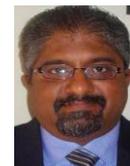